\documentstyle[11pt,moriond,epsfig]{article}

\def\Journal#1#2#3#4{{#1} {\bf #2}, #3 (#4)}

\def\PRD{{\em Phys. Rev.} D}

\def\be{\begin{equation}}
\def\ee{\end{equation}}
\def\bea{\begin{eqnarray}}
\def\eea{\end{eqnarray}}

\def\ar{\rightarrow}
\newcommand{\gsim}{\lower.7ex\hbox{$\;\stackrel{\textstyle>}{\sim}\;$}}
\newcommand{\lsim}{\lower.7ex\hbox{$\;\stackrel{\textstyle<}{\sim}\;$}}

\begin{document}
\vspace*{4cm}
\title{Higgs boson production in association with squark pairs in the
MSSM at the LHC}

\author{ A. Dedes\footnote{Talk given by A. Dedes at the {\it 
XIth Recontres de Blois, ``Frontiers of Matter''}, 
June 27-July 3, 1999, France.} and S. Moretti }

\address{Rutherford Appleton Laboratory,  \\
Chilton, Didcot, Oxon, OX11 0QX, UK}

\maketitle\abstracts{We study neutral and charged Higgs boson production
in association with stop and sbottom
squarks at the Large Hadron Collider (LHC), 
within the so-called M-SUGRA scenario, 
i.e., the Supergravity 
(SUGRA) inspired Minimal
Supersymmetric Standard Model (MSSM). For low values of $\tan\beta$ only
the cases $\tilde{t}_1\tilde{t}_1^* H$, $\tilde{t}_1\tilde{t}_1^* h$ and
          $\tilde{t}_1\tilde{t}_2^* h$            
 give detectable rates while for $\tan\beta \gsim 30$ a variety 
of signals involving all Higgs bosons can be accessed, 
at high collider luminosity. The dependence of these reactions on the M-SUGRA
parameters might further allow one to pin down the actual structure of the 
underlying Supersymmetric (SUSY) model. }

\section{Squark-squark-Higgs boson production at the LHC}
\subsection{Introduction}
We have considered elementary processes of the type
\bea
g+g \ar \tilde{q}_\chi + \tilde{q}^{\prime *}_{\chi^\prime}+\Phi \;,
\label{proc}
\eea
where  $q^{(')}=b,t$, $\chi^{(')}=1,2$ and $\Phi=H,h,A,H^\pm$, in all
possible combinations,  within
the theoretical M-SUGRA framework, 
at the leading order (LO) in QCD. That is, we have
considered $gg$-induced final states involving any possible Higgs 
state produced 
in conjunction with both heavy sbottom and stop squarks.
The phenomenological 
relevance at the LHC of such reactions is in fact twofold 
~\cite{short,large}.

\noindent

$\bullet$ They furnish new  production mechanisms of 
Higgs bosons of the MSSM, both neutral and charged.

$\bullet$  They yield production rates, for particular combinations of 
$q^{(')}$, $\chi^{(')}$ and $\Phi$, strongly dependent
on the fundamental M-SUGRA parameters: i.e., 
$M_0$, $M_{1/2}$, sign($\mu$), $A_0$ and $\tan\beta$.

As for previous literature on the subject, we should mention that the case
 $g g \rightarrow \tilde{t}_1
\tilde{t}_1^* h$ was first
considered in Ref.~\cite{djouadi} in the so-called `decoupling' limit.
Adopting the M-SUGRA scenario, associated production of 
CP-odd Higgs bosons and squark pairs 
was then addressed by the authors
in Ref.~\cite{short}.  Furthermore, in Ref.~\cite{djouadi2}
(see also \cite{fawzi})
light CP-even Higgs boson production in association with light stop squarks was
reanalysed in the M-SUGRA scenario at both Tevatron and
LHC\footnote{The same final state, but produced
in $e^+e^-$ annihilations at the TeV scale,
has been considered in Refs. \cite{djouadi2,ee}.}.
Finally, the study of all final states of the type (\ref{proc}) was completed
 (again, in the M-SUGRA scenario) in Ref.~\cite{large}.
A general consensus on the possible detectability at the LHC of
squark-squark-Higgs events emerged from
Refs.~\cite{short} to \cite{fawzi}, particularly in the case of light 
stop squarks, not too heavy Higgs bosons and/or large trilinear couplings. 

\subsection{Calculation}

Our procedure for calculating the cross sections of the
various processes (\ref{proc}) was the following.

\noindent

$\bullet$  We derived  an analytical expression of the corresponding
$2\to 3$ matrix elements (MEs), which can be found in Ref.~\cite{large}.


$\bullet$  Their integration over the three-body
phase spaces was performed by means of VEGAS~\cite{vegas} and their 
convolution 
with gluon Parton Distribution Functions (PDFs) was provided by the fit 
CTEQ(4L)~\cite{cteq4l},
at the scale $Q=E_{\rm{ecm}}$ (at the partonic level)\footnote{We 
also have resorted to other LO packages, 
such as MRS-LO(09A,10A,01A,07A)~\cite{mrs}: typical differences
were found to be around 15-20\%.}.

$\bullet$ The evolution of the strong coupling constant, $\alpha_s$,
was done at two-loop order with all relevant (s)particle thresholds 
taken into account through the
$\theta$-function approximation~\cite{sakis}.

$\bullet$ In order to prevent the MEs from becoming singular, 
 finite widths were inserted in the 
(anti)squark propagators, as computed from 
the program {\tt ISAJET/ISASUGRA 7.40} \cite{isajet}.

\section{Results}

\subsection{Small $\tan\beta$ region}

In the small region, e.g., $\tan\beta\sim 2$, the following cross sections 
exceed the $1~fb$ level, value which we adopt as threshold 
of observability for a annual luminosity ${\cal L}=100~fb^{-1}$. These
are:
\bea
gg &\ar& \tilde{t}_1\tilde{t}_1^* H, \nonumber \\
gg &\ar& \tilde{t}_1\tilde{t}_1^* h, \nonumber \\
gg &\ar& \tilde{t}_1\tilde{t}_2^* h. \nonumber 
\eea
In fact the largest cross sections are in general those associated
to the last two processes, in decreasing size. These are both strongly 
dependent upon $A_0$, the (common) trilinear coupling of the M-SUGRA model. 
In the case of heavy CP-even Higgs boson production in the
first channel, though the cross sections are typically smaller, 
it is possible to distinguish the sign of the 
 M-SUGRA Higgsino mass, $\mu$, since the  
production rates corresponding to the two opposite signs can differ by up
to two orders of magnitude. As for the other combinations of squarks and
Higgs bosons in (\ref{proc}), their cross sections are always below detection
level.

\subsection{Large $\tan\beta$ region}

If $\tan\beta$ is as large as 30 or more, a larger
variety of  squark-squark-Higgs
processes can be accessed at the LHC, at least for some specific combinations
of M-SUGRA parameters, when both the gaugino and squark masses
are not too heavy at the Plank scale (say, 
$M_0 < 500$ GeV and $M_{1/2} < 220$ GeV) and 
the trilinear couplings are negative, $A_0 < 0$ GeV. 
\begin{figure}
\hbox{\psfig{figure=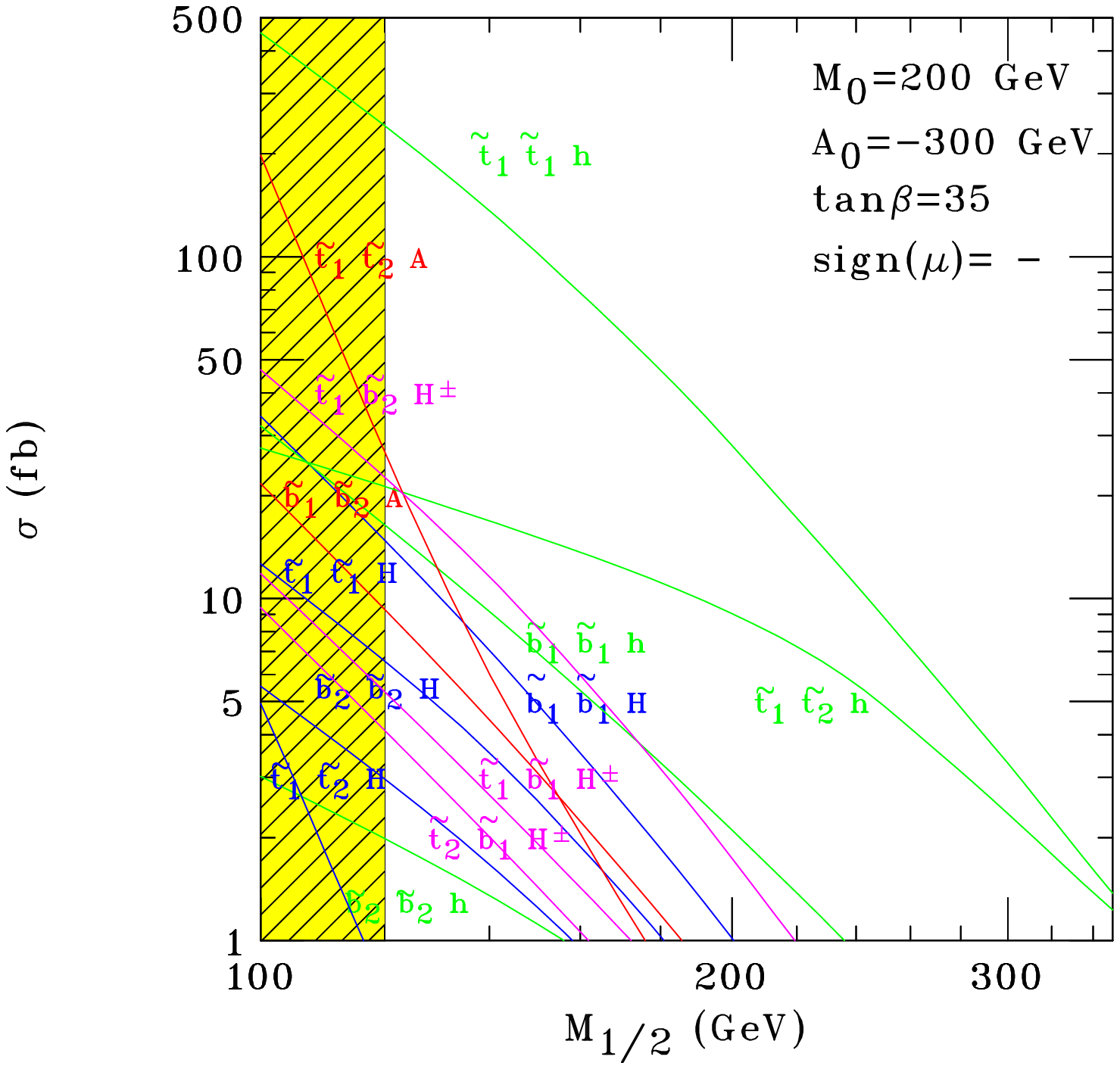,height=3.0in,angle=90}
\psfig{figure=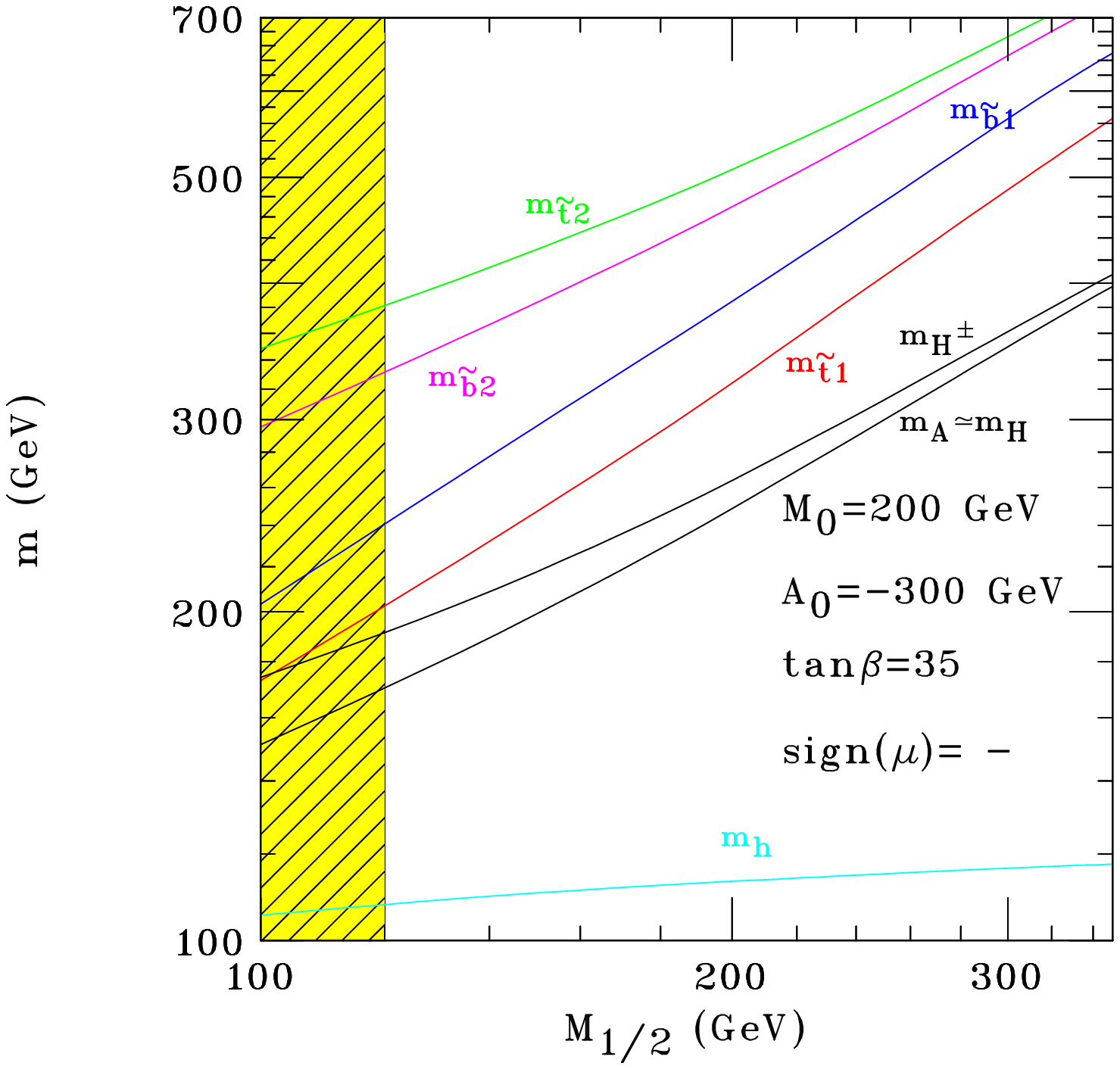,height=3.0in,angle=90}}
\caption{The largest squark-squark-Higgs cross sections at large $\tan\beta$ 
 (left) and the
values of the  masses in the final states  
(right), as a function of $M_{1/2}$.
 Shaded regions indicate areas excluded by direct squark/Higgs
searches. The  choice of the other M-SUGRA parameters is also shown. }
\label{fig}
\end{figure}
These modes are:
\bea
gg &\ar&  \tilde{t}_1 \tilde{t}_1^* H,   \nonumber \\
gg &\ar&  \tilde{b}_1 \tilde{b}_1^* H,   \nonumber \\
gg &\ar&  \tilde{t}_1 \tilde{t}_1^* h,   \nonumber \\
gg &\ar&  \tilde{t}_1 \tilde{t}_2^* H,   \nonumber \\
gg &\ar&  \tilde{b}_1 \tilde{b}_1^* H,   \nonumber \\
gg &\ar&  \tilde{t}_1 \tilde{t}_2^* A,   \nonumber \\
gg &\ar&  \tilde{b}_1 \tilde{b}_2^* H,   \nonumber \\
gg &\ar&  \tilde{t}_1 \tilde{b}_2^* H^-. \nonumber 
\eea
Their production rates are displayed in Fig.~\ref{fig} (left), alongside
the masses of the three final state particles (right).

\subsection{Possible signatures and backgrounds}

Many decay channels of the particles in the final states of
 processes of the type (\ref{proc}) were considered in 
Refs.~\cite{short,large}. Here, we only show one example:
for the case of CP-odd Higgs boson production in association with
stop pairs, i.e., $gg\ar \tilde{t}_1\tilde{t}_2^* A$. The choice of
M-SUGRA parameters is the following:
 $M_0=M_{1/2}=125$ GeV, $A_0=0$ GeV, $\tan\beta=40$ and 
$\mu<0$. One can have the decay pattern:
\begin{center}
\bea
\begin{array}{|c|}
\hline
\sigma(gg\ar \tilde{t}_1\tilde{t}_2^* A) = 96 ~{\rm fb} \\ \hline 
\end{array}
\eea
$\Downarrow$
$$
\arraycolsep=0pt
\begin{array}{lllllll}\label{newchain}
{\tilde{t}_1} && &~~~~{\tilde{t}}_2^* &&A~~~~~ & \\
\downarrow &&& ~~~\downarrow &&\downarrow~~~~~ & \\
\chi_1^+ \,+\,b\; & &&~~~~\chi_1^- \,+\,\bar b &~&b \,+\,\bar b &  \\
\downarrow &&& ~~~\downarrow && \; & \\
{\tilde{\tau}}_1^+ \,+\,\nu\,+\,b& && ~~~~
{\tilde{\tau}}_1^- \,+\,\bar\nu\,+\,\bar b &&& \\
\downarrow &&& ~~~\downarrow && \; & \\
\tau^+~+~\nu\,+\,b\,+\,\chi_1^0 & &&
 ~~~~\tau^-\,+\,\bar\nu~+~\bar b\,+\,\chi^0_1 &&&
\end{array}
$$
$\Downarrow$
\bea
\begin{array}{|c|}\hline
4b + \tau^+\tau^- + E_{miss} \\ \hline
\end{array}
\eea
\end{center}
which leads to 528 detectable events every 100 $fb^{-1}$,  
after multiplying by $\varepsilon_b^4$ (the efficiency to tag four
$b$-quarks, where $\varepsilon_b$ is taken to be 70\%) and accounting for
all relevant branching fractions. Standard backgrounds would come
from `$Z$~+~4$b$' production. However, from one pair of $b$-quarks one could
reconstruct the $A$ mass, at around 114 GeV, thus reducing such a QCD noise.
Besides, the large amount of missing energy, $E_{\mathrm{miss}}$, building up
because of the four neutrinos and two neutralinos, could prove to be
a further good handle against non-SUSY processes.  Here, typical
stop masses are around 380 and 240 GeV, for
${\tilde{t}}_2$ and  ${\tilde{t}}_1$, respectively.
As for irreducible
SUSY backgrounds, notice the poor decay rate
BR$({\tilde{t}}_2^*\ar{\tilde{t}}_1^* Z)\approx7\%$ !

\section{Conclusions}

In summary, we have studied neutral and charged Higgs boson
production in association with all possible combinations of
 stop and sbottom squarks at the LHC, in the
context  of the SUGRA inspired MSSM.
We have found that the cross sections of many of these
processes should be  detectable at high
collider luminosity for not too small values of $\tan\beta$
(except for the case of light and heavy CP-even Higgs boson being produced
in association  
with light stops, which can
be detected also in the very low $\tan\beta$ regime). 
Furthermore, the strong dependence  of these cross sections
on $M_0$, $M_{1/2}$, sign($\mu$), $A_0$ and/or $\tan\beta$,  
would also help to pin down the actual values of these 
fundamental parameters defining the M-SUGRA
scenario~\cite{short,large}. Finally, we have given one representative example
of a possible squark-squark-Higgs
signature and a brief discussion of how to suppress 
the dominant backgrounds. As outlook for the future, we should 
mention the inclusion of 
the CP-violating phases \cite{preparation}, which can strongly affects
the squark-squark-Higgs vertices here involved, as already shown in
Refs.~\cite{phases} for MSSM (neutral) Higgs production via (s)quark loops in
gluon-gluon fusion.

\medskip\noindent
{\underline{\sl Acknowledgments}}~
{\small AD is supported by the Marie Curie Research Training Grant
ERB-FMBI-CT98-3438. 
SM acknowledges financial support from the UK PPARC.
We both thank G. Corcella for illuminating discussions 
during our stay at the Old School, Oxford.}

\end{document}